\documentclass[twocolumn,amsmath,amssymb]{revtex4}

\def\gsim{ \lower .75ex \hbox{$\sim$} \llap{\raise .27ex \hbox{$>$}} }

\def\lsim{ \lower .75ex \hbox{$\sim$} \llap{\raise .27ex \hbox{$<$}} }

\usepackage{graphicx}
\usepackage{dcolumn}
\usepackage{bm}
\input epsf
\def\gsim{ \lower .75ex \hbox{$\sim$} \llap{\raise .27ex \hbox{$>$}} }
\def\lsim{ \lower .75ex \hbox{$\sim$} \llap{\raise .27ex \hbox{$<$}} }

\usepackage{graphicx}
\usepackage{dcolumn}
\usepackage{bm}

\newcommand{\nn}{\nonumber}
\newcommand{\be}{\begin{equation}}
\newcommand{\ee}{\end{equation}}
\newcommand{\bea}{\begin{eqnarray}}
\newcommand{\eea}{\end{eqnarray}}

\def\d{\delta}

\def\s{\sigma}
\def\e{\epsilon}

\def\f{\phi}

\def\k{\kappa}

\def\s{\sigma}

\begin{document}

\title{Intuitive understanding of non-gaussianity in ekpyrotic and cyclic models}

\author{Jean-Luc Lehners$^{1}$ and Paul J. Steinhardt$^{1,2}$}
\affiliation{ $^1$Princeton Center for Theoretical Science,
Princeton
University, Princeton, NJ 08544 USA \\
$^2$Joseph Henry Laboratories, Princeton University, Princeton,
NJ 08544, USA}

\begin{abstract}
It has been pointed out by several groups that ekpyrotic and
cyclic models generate significant non-gaussianity.  In this
paper, we present a physically intuitive, semi-analytic
estimate of the bispectrum.  We show that, in all such models,
there is an intrinsic contribution to the non-gaussianity
parameter $f_{NL}$ that is determined by the geometric mean of
the equation of state $w_{ek}$ during the ekpyrotic phase and
$w_{c}$ during the phase that curvature perturbations are
generated and whose value is ${\cal O}(100)$ or more times the
intrinsic value predicted by simple slow-roll inflationary
models, $f_{NL}^{intrinsic} = {\cal O}(0.1)$.  Other
contributions to $f_{NL}$, which we also estimate, can increase
$|f_{NL}|$ but are unlikely to decrease it significantly,
making non-gaussianity a useful test of these models. Furthermore, we discuss a
predicted correlation between the non-gaussianity and scalar spectral index that sharpens the test.
\end{abstract}

\pacs{PACS number(s): 98.80.Es, 98.80.Cq, 03.70.+k}

\maketitle

\section{Introduction}

Observations of the cosmic microwave background and large scale
structure have provided strong evidence for a nearly
scale-invariant spectrum of primordial adiabatic density
fluctuations with perhaps a slightly red tilt
\cite{Komatsu:2008hk,Seljak:2006bg}. The key cosmological
question is: how was this spectrum generated?  Two mechanisms
are currently known, inflation
\cite{Guth:1982ec,Hawking:1982cz,Starobinsky:1982ee,Bardeen:1983qw}
and ekpyrosis
\cite{Khoury:2001wf,Steinhardt:2001vw,Tolley:2003nx,others,Lehners:2007ac}.
Inflation is a period of accelerated expansion following the
big bang, characterized by a large Hubble parameter $H$ and an
equation of state $w \approx -1$.  Ekpyrosis is a period of
ultra-slow contraction preceding the big bang, characterized by
a small $H$ and $w \gg1$.  One way of distinguishing the two
mechanisms is by measuring the spectrum of primordial
gravitational waves
\cite{Khoury:2001wf,Steinhardt:2001vw,Boyle:2003km} whose
amplitude, proportional to $H^2$, is exponentially different in
the inflationary versus ekpyrotic/cyclic models.  A second
approach, recently emphasized by several groups analyzing
different types of ekpyrotic models \cite{Buchbinder:2007ad,
Creminelli:2007aq,
Koyama:2007if,Buchbinder:2007at,Battefeld:2007st,Lehners:2007wc},
is by measuring the non-gaussian contributions to the density
fluctuation spectrum; {\it e.g.} the bispectrum, parameterized
by the parameter $f_{NL}$ \cite{Komatsu:2000vy}, as well as the
correlation between $f_{NL}$ and the scalar spectral tilt $n_s$
\cite{Lehners:2007wc}.

The purpose of this paper is to provide a physically intuitive
semi-analytic estimate of the contributions to the bispectrum
that makes clear why measuring both the non-gaussianity and the
scalar spectral tilt is a good approach for testing ekpyrotic
scenarios of various types, including the cyclic model. We show
that ekpyrotic and cyclic models generically produce
non-gaussianity of the ``local'' type \cite{Babich:2004gb} with
a net value of $f_{NL}$ that is at least an order of magnitude
greater than the value predicted by simple slow-roll
inflationary models ($f_{NL} \lesssim 1$).  The prediction is
timely because this value is large enough to be detectable in
near-future cosmic microwave background studies.  The current
bound reported by the WMAP collaboration, $-9  < f_{NL} < 111$
and $0.94< n_s < 0.99$ at $2\s$ \cite{Komatsu:2008hk} already
intrudes on the ekpyrotic range. The uncertainty in $f_{NL}$
will improve modestly with further WMAP data, and then
dramatically ($\Delta f_{NL} \le 5$) with the forthcoming
Planck satellite mission and large scale structure studies.

The reason why the non-gaussianity in ekpyrotic/cyclic models
is generically several orders of magnitude greater than in
inflationary models can be traced to the difference in the
equation of state during the period that density fluctuations
are generated.  In standard versions of both models, the
density perturbation spectra have their origin in scalar fields
$\phi_i$ which develop nearly scale invariant perturbations
while evolving along an effective potential $V(\phi_i)$. During
an inflationary phase, the potential must be nearly constant in
order to obtain $w_{inf} \approx -1$ or, equivalently,
$\epsilon_{inf} \equiv \frac{3}{2} (1+w_{inf}) \ll 1.$ This
means that the inflaton is a nearly free field with nearly
gaussian quantum fluctuations. The non-gaussian amplitude
depends on the deviation of the potential from perfect flatness
or, equivalently, on how close the slow-roll parameter
$\epsilon_{inf}$ and its time variation are to zero. This
intuitive argument is consistent with the quantitative
expression obtained by Maldacena \cite{Maldacena:2002vr}, for
example. By contrast, a  negative, exponentially steep
potential is required to obtain $w_{ek}\gg1$ or, equivalently,
 $\epsilon_{ek} \gg 1$ in an
ekpyrotic phase, which means that the scalar fields have
significant nonlinear self-interactions whose magnitude depends
on how large $\epsilon_{ek}$ is. Because the magnitude of
$\epsilon_{ek}$ is ${\cal O}(100)$ or more times larger than
$\epsilon_{inf}$, the scalar field contribution to the
non-gaussianity -- which we will call the ``intrinsic'' part --
is significantly larger for ekpyrotic/cyclic models
\cite{others,Buchbinder:2007ad, Creminelli:2007aq,
Koyama:2007if,Battefeld:2007st,Lehners:2007wc}.

This intuitive argument only refers to the intrinsic
contribution to the non-gaussianity, but this is enough to
argue why ekpyrotic/cyclic models generically predict
$|f_{NL}|$ to be several orders of magnitude greater than the
value in inflationary models. Even if the additional effects
discussed below add or subtract from $|f_{NL}|$ in the
ekpyrotic/cyclic model, obtaining a value less than one (that
is, in the inflationary range) would only occur through
accidental cancellations of independent terms at the two- or
three-decimal level, which is highly unnatural
\cite{TurokException}. Conversely, it is possible to add
features to inflationary models (such as curvatons,
non-standard kinetic energy density, etc., with certain
parameters) that enhance the non-gaussianity beyond
$|f_{NL}|=1$.  However, these are unnecessary embellishments,
and, when added, can produce virtually arbitrary $f_{NL}$ of
either sign.

The predictions can be further refined by taking account of how
the scalar field fluctuations are transformed into curvature
perturbations. In inflationary models, the scalar field
fluctuations directly produce curvature perturbations that are
growing modes in an expanding universe. Consequently, the tiny
intrinsic non-gaussianity in the scalar fields discussed above
translates directly to a tiny non-gaussianity in the curvature
fluctuations (if no embellishments are added).  In
ekpyrotic/cyclic models, the curvature perturbations produced
directly by the scalar fields are decaying modes. Hence, the
process involves scalar field fluctuations first producing
growing mode entropic perturbations  during the ekpyrotic phase
and, then, converting them to curvature perturbations just
before the bounce to an expanding phase
\cite{others,Lehners:2007ac}. It is notable that the equation
of state during this conversion (or $\epsilon_{c}$) can be
quite different from the equation of state during the ekpyrotic
phase (or $\epsilon_{ek}$). Since the entire curvature
perturbation is produced by this conversion, even the intrinsic
contribution is necessarily affected by $\epsilon_{c}$, as well
as $\epsilon_{ek}$.  In fact, as we will show below, the
intrinsic $f_{NL}$ turns out to be proportional to their
geometric mean. Hence, the magnitude of $f_{NL}$ in
ekpyrotic/cyclic models can be considerably less if the
conversion takes places in a kinetic energy dominated phase
when $\epsilon_{c} = 3 \ll \epsilon_{ek}$,  say, rather than in
the ekpyrotic phase \cite{Lehners:2007wc}. This accounts for
why the predictions for $f_{NL}$ were found to be significantly
greater in the new ekpyrotic model \cite{Buchbinder:2007ad,
Creminelli:2007aq, Koyama:2007if} compared to the cyclic model
\cite{Lehners:2007wc}, although both models predict values much
greater than the inflationary case.

The conversion mechanism is also important for determining the
sign of the intrinsic contribution to $f_{NL}$.  For example,
the cyclic model discussed in Ref.~\cite{Lehners:2007wc}
produces either sign for the intrinsic $f_{NL}$, whereas the
sign is typically negative for the case considered in
Ref.~\cite{Buchbinder:2007ad, Creminelli:2007aq,
Koyama:2007if}.
 Furthermore, the conversion from
entropic to curvature perturbations necessarily introduces an
additional contribution to the non-gaussianity, which will be
analyzed below.  We will see that there are substantial regimes
where this contribution is smaller in magnitude compared to the
intrinsic contribution, but also substantial regimes where it
is larger and can even reverse the sign. In the latter case,
the net non-gaussianity tends to be so large that it is already
ruled out by existing experiments.

This paper is designed to translate the intuitive discussion
above into simple semi-analytic estimates, making clear those
aspects that are highly model dependent and those that are
generic. The conclusions themselves are not so original; for
the most part, they appear in earlier papers focusing on
particular examples
\cite{others,Buchbinder:2007ad,Creminelli:2007aq,Koyama:2007if,Battefeld:2007st,
Lehners:2007wc}. Our intent here is more modest: to provide
simple expressions that clarify their origin and physical
interpretation, so that the significance of forthcoming
non-gaussianity tests can be better appreciated. The
organization of the paper is as follows: in the next section we
set up our notation while reviewing the context of the
calculations performed later in the paper. In section
\ref{section intrinsic} we derive the intrinsic non-gaussianity
present in the entropy perturbation in ekpyrotic models. This
allows us to present an order-of-magnitude estimate of the
resulting curvature perturbation in section \ref{section
conversion}, before proceeding to discuss various conversion
mechanisms in more detail. Section \ref{section conclusion}
contains the conclusions that we draw from our analysis.

\section{Setup} \label{section setup}

In the ekpyrotic/cyclic model, the 4d effective theory
describing the universe around the time of the big bang is
given by  gravity coupled to two minimally coupled scalar
fields with potentials. (This is the case in heterotic M-theory
\cite{Lehners:2006ir}, for example.): \be \int \d^4 x \sqrt{-g}
\,\Big(\, \frac{1}{2}R -\frac{1}{2} \sum_{i=1}^{2} (\partial
\phi_{i})^2 -\sum_{i=1}^{2} V_i(\phi_i)\,\Big). \label{lag} \ee
The scalar field and Friedmann equations are given by \be
\ddot{\phi}_i + 3H\dot{\phi}_i + V_{i,\phi_i} = 0 \ee and \be
H^2 =\frac{1}{3} \left[\frac{1}{2}  \sum_i \dot\phi_i^{~2}+
\sum_i V_i(\phi_i)
 \right], \ee where $H=\dot{a}/a,$ $a$ denotes the scale factor, and
$V_{i,\phi_i} = (\partial V_i/\partial \phi_i)$ with no
summation implied.  In a contracting universe, a growing mode
is given by the entropy perturbation, namely the relative
fluctuation in the two fields, defined (at linear order) as
follows \cite{Gordon:2000hv} \be \delta s \equiv
(\dot{\phi}_1\,\delta \phi_2 - \dot{\phi}_2\,\delta
\phi_1)/\dot{\s}, \ee where we have defined \be \dot{\s} \equiv
\sqrt{\dot{\phi}_1^2+\dot{\phi}_2^2}. \label{c3} \ee The
entropy perturbation is gauge-invariant and it represents the
perturbation orthogonal to the background scalar field
trajectory, as shown in Fig.~\ref{FigureSubdominant}; see
Ref.~\cite{Langlois:2006vv} for its definition to all orders.
Its equation of motion, up to second order in field
perturbations, is \cite{Langlois:2006vv} \bea && \ddot{\delta
s} +3H \dot{\delta s}+ \left(V_{ss}+3\dot{\theta}^2 \right)
\delta s \nn
\\ && +\frac{\dot{\theta}}{\dot{\sigma}}(\dot{\delta s}^{(1)})^2  +\frac{2}{\dot{\sigma}}\left( \ddot{\theta}+
\dot{\theta}\frac{V_{\sigma}}{\dot{\sigma}} -
\frac{3}{2}H\dot{\theta}\right)\delta s^{(1)} \dot{\delta
s}^{(1)} \nn
\\ && +\left( \frac{1}{2}
V_{sss}-5\frac{\dot{\theta}}{\dot{\sigma}}V_{ss}-
9\frac{\dot{\theta}^3}{\dot{\sigma}} \right)(\delta s^{(1)})^2
= 0, \label{eq-entropy} \eea where we have neglected a
non-local term that is unimportant in ekpyrotic models
\cite{Lehners:2007wc}. Note that this is a closed equation for
the entropy perturbation. Here $V_{\sigma}$ denotes a
derivative of the potential along the background trajectory and
the successive derivatives of the potential with respect to the
entropy field are given by \bea V_{s} &=&
\frac{1}{\dot{\s}}(\dot{\phi_1} V_{,\phi_2}- \dot{\phi_2}
V_{,\phi_1})\\ V_{ss} &=& \frac{1}{\dot{\s}^2} (\dot{\phi_1}^2
V_{,\phi_2 \phi_2}-2 \dot{\phi_1}\dot{\phi_2}
V_{,\phi_1 \phi_2} + \dot{\phi_2}^2 V_{,\phi_1 \phi_1}) \\
V_{sss} &=& \frac{1}{\dot{\s}^3}(\dot{\phi_1}^3 V_{,\phi_2
\phi_2
\phi_2}-3\dot{\phi_1}^2\dot{\phi_2} V_{,\phi_1 \phi_2 \phi_2} \nn \\
& & + 3\dot{\phi_1}\dot{\phi_2}^2 V_{,\phi_1 \phi_1 \phi_2} -
\dot{\phi_2}^3 V_{,\phi_1 \phi_1 \phi_1}). \eea The angle
$\theta$ of the background trajectory is defined by
\cite{Gordon:2000hv} $ \cos(\theta)=\dot{\f}_1/\dot{\s} \, , \,
\sin(\theta) = \dot{\f}_2/\dot{\s}$ and thus $\dot{\theta} =
-V_{s}/\dot{\s}.$ The  parameter $\epsilon$ and the equation of
state parameter $w$ are related to the background evolution via
\be \epsilon \equiv \frac{3}{2} (1+w) \equiv
\frac{\dot{\sigma}^2}{2H^2}, \ee  and the ekpyrotic phase can
be characterized by $\epsilon \gg 1.$

\section{Intrinsic non-gaussianity} \label{section intrinsic}

The entropy perturbations are generated during the ekpyrotic
phase, when the potentials are given by \be \sum_{i=1}^{2}
V_i(\phi_i)\,= -V_1 e^{-\int c_1 \d \phi_1}-V_2 e^{-\int c_2
\d\phi_2}, \label{c1} \ee {\it i.e.} the potentials are
negative and steep. Here we consider $c_1=c_1(\phi_1)>0$ and
$c_2 = c_2(\phi_2)$ to be slowly varying in time, while $V_1$
and $V_2$ are positive constants. The ekpyrotic phase quickly
flattens the universe, so that we can assume a flat
Friedmann-Robertson-Walker (FRW) background with line element
$\d s^2= -\d t^2 +a^2(t) \d {\bf x}^2.$ If the $c_i$ are
exactly constant and moreover $|c_i| \gg 1,$ then the
Einstein-scalar equations admit the scaling solution \be a =
(-t)^{1/\e}, \quad \phi_i = {2\over c_i} \ln
(-\sqrt{c_i^2V_i/2} t), \quad \frac{1}{\e}= \sum_i {2 \over
c_i^2}. \label{c2} \ee This solution describes a very slowly
contracting universe with $\e \gg 1$. During the  phase in
which the entropic perturbations are generated,
$\dot{\theta}=0$, which corresponds to a straight background
scalar field trajectory. In this case, we define \be
\dot{\phi}_2 \equiv \gamma \dot{\phi}_1, \label{phirel} \ee
and, with this notation, $c_1 = \gamma c_2$ and
 \be \epsilon_{ek}=
\frac{|\gamma  c_1 c_2| }{2(1+\gamma^2)}. \ee Results are not
very sensitive to $\gamma$ so we take
 $\gamma = {\cal O}(1)$ throughout.

It is useful to recast the evolution in terms of the adiabatic
and entropic variables $\s$ and $s$. Up to unimportant additive
constants (that we will fix below), they can be defined by \bea
\s &\equiv& \frac{\dot\phi_1 \phi_1 + \dot\phi_2
\phi_2}{\dot\s}
\\ s &\equiv& \frac{\dot\phi_1 \phi_2 - \dot\phi_2
\phi_1}{\dot\s}. \eea Then we can expand the potential up to third order as follows
\cite{Buchbinder:2007tw}: \be V_{ek}=-V_0 e^{\sqrt{2\e}\s}[1+\e
s^2+\frac{\k_3}{3!}\e^{3/2} s^3],
\label{potentialParameterized}\ee where $\k_3$ is of ${\cal O}(1)$
 for typical potentials (the case of exact exponentials corresponds to $\k_3=-4\sqrt{2/3}$).
The scaling
solution (\ref{c2}) can be rewritten as \be a(t)=(-t)^{1/\e} \qquad
\s=-\sqrt{\frac{2}{\e}}\ln \left(-\sqrt{\e V_0} t\right) \qquad
s=0. \label{ScalingSolution}\ee  The intrinsic non-gaussianity
in the entropy perturbation is produced during the ekpyrotic
phase and can be determined from the equation of motion for the
entropy field (\ref{eq-entropy}), which reduces to \be
\ddot{\delta s}+3H\dot{\delta s}+V_{ss}\delta s +
\frac{1}{2}V_{sss} (\delta s^{(1)})^2 = 0.
\label{EomGeneration}\ee The last term is \be
\frac{1}{2}V_{sss} =-\frac{\k_3}{2t^2} \sqrt{\e}
 \ee and thus the solution, at long wavelengths and up to second order
in field perturbations, is given by
\cite{Koyama:2007if,Lehners:2007wc} \bea \delta s(t) &=& \delta
s^{(1)}(t) + \delta s^{(2)}(t) \nn
\\ &=& \delta s_{end} \frac{t_{end}}{t} + \tilde{c} (\delta
s_{end} \frac{t_{end}}{t})^2, \label{entropys2ekpyrosis}\eea
where \be \tilde{c} = \frac{\k_3\sqrt{\e}}{8}.
\label{ctildedefinition}\ee Thus, the intrinsic non-gaussianity
present in ekpyrotic models (second term in
(\ref{entropys2ekpyrosis}) above) is of order ${\cal{O}}
(\sqrt{\epsilon_{ek}})$ and it is due to the steepness of the
potentials and the resulting self-interactions of the scalar
fields. This is in sharp contrast with inflation, where the
intrinsic non-gaussianity is extremely small due to the
flatness of the potential (or, equivalently, $\epsilon_{inf}\ll
1$).

\section{Conversion} \label{section conversion}

\subsection{An estimate of intrinsic $f_{NL}$} \label{subsection
estimate}

What is measured is not directly the non-gaussianity present in
the entropy perturbation, but the non-gaussianity it imprints
on the curvature perturbation. Thus, it is important to know
the strength with which this intrinsic non-gaussianity gets
transferred to the curvature perturbation. The time evolution
of the curvature perturbation to second order in field
perturbations and at long wavelengths is given in FRW time by
\cite{Langlois:2006vv} \be \dot{\cal{R}} =
\frac{2H}{\dot{\s}}\dot{\theta}\delta s +
\frac{H}{\dot{\s}^2}[- (V_{ss} + 4 \dot{\theta}^2) (\delta
s^{(1)})^2+\frac{V_{,\sigma}}{\dot\s}\d s \dot{\d s}].
\label{zetadotquadratic} \ee
where we have omitted a non-local term which can be neglected
in ekpyrotic models for the same reason as the non-local term
omitted from (\ref{eq-entropy}) above, see
\cite{Lehners:2007wc} for details.
At linear order, a non-zero entropy perturbation combined with
a bending ($\dot{\theta}\neq 0$) of the background trajectory
sources the curvature perturbation on large scales and results
in a linear, gaussian curvature perturbation \bea {\cal{R}}_L
&=& \int_{\Delta t} \frac{2H}{\dot{\sigma}}\dot{\theta}\delta
s^{(1)} \\ &=& \int_{\Delta t} -
\sqrt{\frac{2}{\epsilon_c}}\dot{\theta}\delta s^{(1)},
\label{curvaturelinear} \eea where we denote the duration of
the conversion by $\Delta t$ and the sign corresponds to a
contracting universe. Thus the strength of conversion is
proportional to $1/\sqrt{\epsilon_c}$ and this dependence on
the equation of state will have repercussions for the magnitude
of the second order correction as well.

Here, as with all the contributions to non-gaussianity
discussed in this paper, the fluctuations are generated by the
scalar fields with canonical kinetic energy density, so they
generate non-gaussianity of the ``local'' type, as defined in
Refs.~\cite{Babich:2004gb,Komatsu:2000vy,Maldacena:2002vr}. The
local wavelength-independent non-gaussian contribution to
${\cal R}$ can then be characterized in terms of the leading
linear, gaussian curvature perturbation ${\cal R}_{L}$
according to \be
{\cal{R}}={\cal{R}}_L-\frac{3}{5}f_{NL}{\cal{R}}_L^{2}, \ee
using the sign convention for wavelength-independent
non-gaussianity parameter $f_{NL}$
 in \cite{Komatsu:2000vy}.
Then the contributions to $f_{NL}$ can be divided into three
parts \cite{Lehners:2007wc}  \bea
f_{NL}^{intrinsic}&=&-\frac{5}{3{\cal{R}}_L^2}\int_{\Delta t}
\frac{2H}{\dot{\s}}\dot{\theta}\delta s^{(2)} \label{fNLintrinsic}\\
f_{NL}^{reflection}&=&\frac{5}{3{\cal{R}}_L^2}\int_{ref}
\frac{H}{\dot{\s}^2}[(V_{ss} + 4 \dot{\theta}^2) (\delta
s^{(1)})^2 \nn \\ && -\frac{V_{,\s}}{\dot\s}\d s\dot{\d s}] \label{fNLreflection}\\
f_{NL}^{integrated}&=&\frac{5}{12{\cal{R}}_L^2}(\delta
s^{(1)}(t_{end}))^2. \label{fNLintegrated}\eea
$f_{NL}^{intrinsic}$ arises from the direct translation of the
intrinsic non-linearity present in the entropy perturbation
into a corresponding non-linearity in the curvature
perturbation. By contrast, $f_{NL}^{reflection}$ and
$f_{NL}^{integrated}$ would be non-zero even if the entropy
perturbation were exactly gaussian. Both contributions are due
to the non-linear relationship between the curvature
perturbation and the entropy perturbation, as expressed in
equation (\ref{zetadotquadratic}) - the difference is that
$f_{NL}^{integrated}$ gets generated during the ekpyrotic phase
and $f_{NL}^{reflection}$ during the conversion.

The intrinsic non-gaussianity can be estimated by combining
equations (\ref{entropys2ekpyrosis}), (\ref{curvaturelinear})
and (\ref{fNLintrinsic}):\bea f_{NL}^{intrinsic} &\approx& \pm
\frac{5}{3{\cal{R}}_L^2}\int_{\Delta
t} \sqrt{\frac{2}{\epsilon_c}} \dot{\theta} \tilde{c}{\delta s^{(1)}}^2 \nn \\
&\sim& \sqrt{\epsilon_c \epsilon_{ek}}. \label{fNLestimate}
\eea It is given by the geometric mean of the $\epsilon$
parameters during the phases of generation and conversion of
the perturbations. This is perhaps the most important result
because it fixes a rough magnitude for $|f_{NL}|$ which can
only be significantly reduced through accidental cancellations
due to the  other terms or other effects not included in the
present analysis. It explains in a nutshell why the
non-gaussianity in ekpyrotic/cyclic models is necessarily more
than an order of magnitude greater than in simple inflation
models, and it explains why the equation of state during
conversion can have a significant quantitative effect on the
prediction.

To go beyond this qualitative estimate to a more precise one,
we need to take account of the details of the conversion
mechanism. We will now discuss the various possibilities
considered in the literature, namely conversion during the
ekpyrotic phase, after the ekpyrotic phase during kinetic
energy domination, and conversion after the big bang by
modulated preheating.

\subsection{Conversion during kinetic energy domination}
\label{subsection kinetic}

In the original ekpyrotic and cyclic models,
 the phase dominated by the steep,
 ekpyrotic potential $V(\phi)$ comes to an end
(at $t=t_{end}<0$) before the big crunch/big bang transition
($t=0$), and the universe becomes dominated by the kinetic
energy of the scalar fields. Consequently, the equation of
state at $t=t_{end}$ changes from $\epsilon_{ek} \gg 1$ to
$\epsilon = 3$ (corresponding to $w \rightarrow 1$, the
equation of state for a kinetic energy dominated universe). In
this subsection, we consider the case where the conversion from
entropic to curvature perturbations occurs during this kinetic
energy dominated phase. This occurs naturally in the heterotic
M-theory embedding of the cyclic model because  the
negative-tension brane bounces off a spacetime singularity
\cite{Lehners:2007nb}  --  creating a bend in the trajectory in
field space in the 4d effective theory -- before it collides
with the positive-tension brane (the big crunch-big bang
transition). This bending of the trajectory automatically
induces the conversion of entropy to curvature perturbations
\cite{Lehners:2007ac}. Here we will not restrict the analysis
to this particular example, but we will consider the general
situation in which the scalar field trajectory bends in a
smooth way during the phase of kinetic energy domination
following an ekpyrotic phase.
Before proceeding, we should remark that due to the instability
of the scaling solution during the ekpyrotic phase
\cite{Lehners:2007ac,Tolley:2007nq}, we must assume the
background trajectory is localized very close to the ridge of
the two-field ekpyrotic potential at the beginning of the
ekpyrotic phase in order for the conversion not to happen
already during the ekpyrotic phase. How this initial condition
can arise in a cyclic model is beyond the scope of the present
paper.

We can immediately perform an order-of-magnitude estimate of
the non-gaussianity since during the kinetic phase
$\epsilon_c=3,$ and so (\ref{fNLestimate}) would lead us to
expect $f_{NL}$ to be of order \be f_{NL} \sim
\sqrt{\epsilon_{ek}} \sim {\cal{O}}(c_1). \ee

\begin{figure}[t]
\begin{center}
\includegraphics[width=0.45\textwidth]{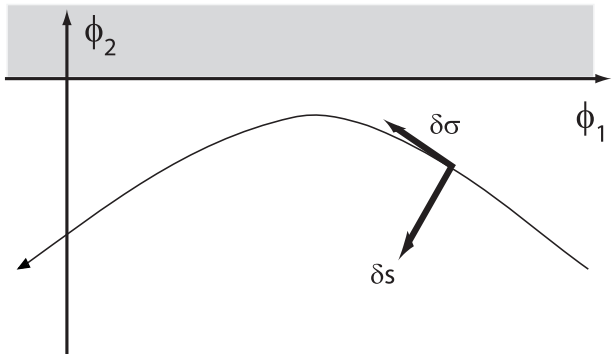}
\caption{\label{FigureSubdominant} {\small The trajectory in field
space reflects off a boundary at $\phi_2=0.$ The entropy
perturbation, denoted $\delta s$, is orthogonal to the trajectory.
The bending causes the conversion of entropy modes into adiabatic modes $\delta \sigma$,
which are perturbations tangential to the trajectory. }}
\end{center}
\end{figure}

We will treat the bend as if only one field reflects
($\phi_2$), and we will consider the case where $\dot\theta >
0,$ see Fig. \ref{FigureSubdominant}. Other cases can be
related to these representative examples by changing the
coordinate system in field space appropriately. In the
heterotic M-theory example, the reflection occurs because
$\phi_2$ comes close to a boundary of moduli space ($\phi_2=0$)
and is forced to bounce \cite{Lehners:2007nb}. For the purposes
of this study, we will treat the reflection as being due to a
potential, $V^R(\phi_2),$ an additional contribution unrelated
to the exponential potentials that were dominant during the
ekpyrotic phase (but which are negligible during the kinetic
energy dominated phase).

It is important to know the evolution of the entropy
perturbation during the process of conversion. If there is a
phase of pure kinetic energy domination before the conversion,
then the background scalar field trajectory is also a straight
line during this phase, but with the potentials being
irrelevant. The equation of motion reduces to \be \ddot{\delta
s}+\frac{1}{t}\dot{\delta s} = 0, \label{EomKE}\ee and by
matching onto the ekpyrotic solution and its first time
derivative at $t=t_{end}$ we find \bea
\delta s(t) &=& \delta s^{(1)}(t) + \delta s^{(2)}(t) \nn \\
&=& \delta s_{end}(1+ \ln{\frac{ t_{end}}{t}}) + \tilde{c}
\delta s_{end}^2 (1+2 \ln{\frac{t_{end}}{t}}). \eea
Incidentally, note that \be t\frac{ \dot{\delta s}}{\delta s}
\sim \frac{1}{\ln(-t)} \label{c4} \ee during the kinetic phase,
while during the ekpyrotic phase $t\dot{\delta s} \sim \delta
s.$ This observation will simplify our analysis later on.

\begin{figure}[t]
\begin{center}
\includegraphics[width=0.45\textwidth]{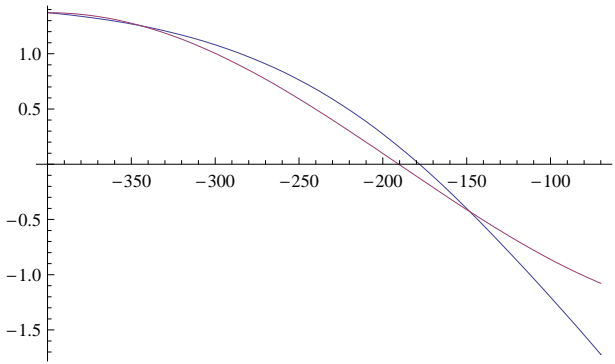}
\caption{\label{Figure3} {\small
The evolution of the linear entropy perturbation during conversion: the solid
(blue)
line shows the actual evolution calculated numerically, while the dashed
(purple) line
shows the approximate solution (\ref{entropys1conversion}) with $\omega=3,$
$\dot{\theta}=-1/t_{ref}$ and $t_{ref} =-400 M_{Pl}^{-1}$. For the
purposes of illustration, $\delta
s^{(1)}(t_{ref})$ has  been normalized to $1$.}}
\end{center}
\end{figure}

During the conversion, even though the kinetic energy of the
scalar fields is still the dominant contribution to the total
energy, the potential $V^R(\phi_2)$ that causes the bending has
a significant influence on the evolution of the entropy
perturbation. To analyze this, we will approximate (as in
\cite{Buchbinder:2007at}) the bending of the trajectory to be
gradual by taking $\dot{\theta}$ constant and non-zero for a
period of time $\Delta t,$ starting from $t=t_{ref}.$ Note
that, assuming the total angle of bending is ${\cal O}(1)$
radian, we have $|\dot{\theta}| \approx 1/|t_{ref}|$ in this
case. Then one can relate the derivatives of the potential to
expressions involving $\dot{\theta},$ for example \be
V^R_{ss}=-2\dot{\theta}^2+\frac{\dot{\theta}}{\gamma
t}.\label{Vssapproximation} \ee Assuming a gradual conversion
($\Delta t \sim t_{ref}$), we can ignore higher derivatives of
$\theta$. (In \cite{Lehners:2007wc} it was shown that sharp
transitions lead to unacceptably large values of $f_{NL};$
hence these cases are of less phenomenological interest.) To
satisfy the constraint on the amplitude of the curvature
perturbation obtained by the WMAP observations, we set $t_{ref}
\approx - 10^3 M_{Pl}^{-1}$ \cite{Lehners:2007ac}. At linear
order, the equation of motion (\ref{eq-entropy}) then reads \be
\ddot{\delta s}^{(1)}+3H\dot{\delta
s}^{(1)}+(\dot{\theta}^2+\frac{1}{\gamma t}\dot{\theta})\delta
s^{(1)} = 0, \label{EomReflectionLinear}\ee As indicated by
equation (\ref{c4}), we can set $\dot{\delta s}^{(1)} = 0$ as a
first approximation and, thus,
 neglect the damping term in
the equation of motion. Also, we will simply evaluate the
coefficient of the last term midway through the reflection, and
define \be \omega \equiv \dot{\theta}
\sqrt{1+\frac{1}{\dot{\theta}\gamma (t_{ref}+\Delta t/2)}}. \ee
For a gradual reflection $\omega \approx (2-3) \dot{\theta}.$
Then the solution for the linear entropy perturbation is \be
\delta s^{(1)} = \delta
s^{(1)}(t_{ref})\cos{\omega(t-t_{ref})}.
\label{entropys1conversion}\ee Instead of continuing to grow
logarithmically, the entropy perturbation actually falls off
during the conversion, see Fig.~\ref{Figure3}. This has the
consequence that the conversion from entropy to curvature
perturbations is less efficient than one might have naively
thought. From (\ref{curvaturelinear}), we can estimate \be
{\cal{R}}_L= -\sqrt{\frac{2}{3}}\dot{\theta}\int_{\Delta t}
\delta s^{(1)}= -\sqrt{\frac{2}{3}} \frac{\dot{\theta}}{\omega}
\delta s^{(1)}(t_{ref})\sin{\omega \Delta t},
\label{curvaturelinestimate}\ee where we have used
$\epsilon_c\approx 3$,  which is a  good approximation for
subdominant reflections (and acceptable for the estimating
purposes for dominant ones).
Since $\omega \Delta t \approx 3,$ the entropy perturbation
evolves over nearly a half-cycle and consequently $\sin \omega
\Delta t$ is a small factor which we will take to be about
$1/3$ in our estimates, which fits well with numerical results.

We also need to know the evolution of the second order entropy
perturbation during the time of conversion. Using the same
approximations as above (implying that we can neglect $V_{sss}$
compared to $\dot\theta V_{ss}/\dot\sigma$ and
$\dot\theta^3/\dot\sigma$ by the time the conversion is
underway), the equation of motion (\ref{eq-entropy}) simplifies
to \be \ddot{\delta s}^{(2)}+ \omega^2 \delta s^{(2)} -
\frac{\dot{\theta}}{\dot{\sigma}} \omega^2 (\delta s^{(1)})^2 =
0, \label{entropys2reflection} \ee where $\delta s^{(1)}$ is
given in (\ref{entropys1conversion}). Putting
$\dot{\theta}\approx \dot{\sigma}$ at the start of the
reflection and keeping in mind that we impose the boundary
condition $\dot{\delta s}^{(2)}\approx 0$, the solution for the
second order entropy perturbation is given by \bea  \delta
s^{(2)} &=& \delta
s^{(2)}(t_{ref})\cos[\omega(t-t_{ref})] \nn \\
& & +\frac{1}{12}(\delta s^{(1)}(t_{ref}))^2 \Big(
-4\cos[\omega(t-t_{ref})] \nn \\ && +4\cos^4[\omega(t-t_{ref})]
+9\sin^2[\omega(t-t_{ref})] \nn
\\ & & +\sin[\omega(t-t_{ref})]\sin[3\omega(t-t_{ref})] \Big). \label{s2}
\eea At large $\epsilon_{ek}$ (or, equivalently, large
$|\tilde{c}|$), the second order entropy perturbation falls off
in the same way as the linear perturbation, but at small
$|\tilde{c}|$ there are significant corrections to this
behavior. We will also need the integral \bea C^{-2}
\int_{\Delta t}\delta s^{(2)}   &=&
\frac{(1+2\ln(t_{end}/t_{ref}))}{(1+\ln(t_{end}/t_{ref}))^2}
\frac{\sin(\omega \Delta t)}{ \omega}\tilde{c} \nn \\
&& +\frac{\Delta t}{2} -\frac{\sin(\omega \Delta
t)}{3\omega}-\frac{\sin(2\omega \Delta t)}{12\omega}, \eea
where $C= \delta s^{(1)}(t_{ref})$. In all cases of interest,
the last two terms are negligible.

We are finally in a position to evaluate the various
contributions to the non-linearity parameter $f_{NL}.$ The
intrinsic contribution, defined in (\ref{fNLintrinsic}),
becomes \bea f_{NL}^{intrinsic} \approx A \k_3
\sqrt{\epsilon_{ek}}+B, \label{simp} \eea with \bea A &=&
\frac{5\omega(1+2\ln{(t_{end}/t_{ref})})}{8\sqrt{6}\dot{\theta}
(1+\ln{(t_{end}/t_{ref})})^2\sin(\omega \Delta t)} \nn \\ B &=&
\frac{5\omega^2}{2\sqrt{6}\dot{\theta}^2\sin^2(\omega \Delta
t)}. \eea Eq.~(\ref{simp}) is one of our key results because it
shows that the essential contribution scales in a simple way
with $\epsilon_{ek}$ and has a value that exceeds the total
$f_{NL}$ for simple inflationary models by more than an
 order of magnitude.
As suggested by Eq.~(\ref{fNLestimate}), the first term in
$f_{NL}^{intrinsic}$ can be re-expressed as ${\cal O}
\,(\sqrt{\epsilon_c \epsilon_{ek}})$ with $\epsilon_c=3$. This
contribution comes directly from the non-zero $\delta s^{(2)}$
generated during the ekpyrotic phase, which is due to the third
derivative w.r.t. $s$ of the ekpyrotic potential.  It is
interesting to note that, in the case where $V_{sss}=0$ or
$\k_3=0$, $f_{NL}^{intrinsic}$ is nevertheless non-negligible
because of the positive offset $B$ generated by the linear
entropy perturbation  $\delta s^{(1)}$ during conversion, the
piece proportional to  $(\delta s^{(1)})^2$ in (\ref{s2}).

Using the definition (\ref{fNLreflection}) together with
(\ref{Vssapproximation}) it is straightforward to see that for
conversion during the kinetic phase, $f_{NL}^{reflection}$ is
always negative, and it can be estimated as \bea
f_{NL}^{reflection} &\approx&
\frac{5}{6{\cal{R}}_L^2}\int_{\Delta t} (2 \dot{\theta}^2 t +
\frac{\dot{\theta}}{\gamma}) (\delta s^{(1)})^2 \nn \\
&\approx& -\frac{15\omega^2}{8\dot{\theta}^2\sin^2(\omega
\Delta t)}\frac{|t_{ref}+\Delta t/2|}{\Delta t}, \eea  where we
have used $\gamma t_{ref} \dot{\theta} \approx 1$ and \be
\int_{\Delta t} t \sin^2 \omega (t-t_{ref})\approx
\frac{1}{2}(t_{ref}+\frac{\Delta t}{2})\Delta t, \ee which is a
valid approximation since $\delta s^{(1)}$ evolves over
approximately a half-cycle. We have neglected the contribution
due to the term proportional to $\d s \dot{\d s},$ since it is
suppressed on account of (\ref{c4}). The integrated
contribution to $f_{NL}$ generated during the ekpyrotic phase
gives an additional contribution of
 \bea
f_{NL}^{integrated}&=&\frac{5}{12 {\cal{R}}_L^2}(\delta
s^{(1)}(t_{end}))^2 \nn \\ &\approx&
\frac{5\omega^2}{8\dot{\theta}^2(1+\ln{(t_{end}/t_{ref})})^2\sin^2(\omega
\Delta t)}. \eea Note that neither the reflected nor the
integrated contributions depend on $\epsilon_{ek};$ they both
simply shift the final result by a number depending on the
sharpness of the transition and the duration of the purely
kinetic phase respectively.

The total $f_{NL}$ is the sum of all the above contributions.
Since we are only considering gradual conversions, we expect
the ratio $|t_{ref}+\Delta t/2| /\Delta t$ to lie between 1 and
2. Also, the duration of the pure kinetic phase between the end
of the ekpyrotic phase and the conversion is necessarily rather
short, so that we expect $\ln(t_{end}/t_{ref}) \lesssim {\cal
O}(1).$

\begin{figure}[t]
\begin{center}
\includegraphics[width=0.45\textwidth]{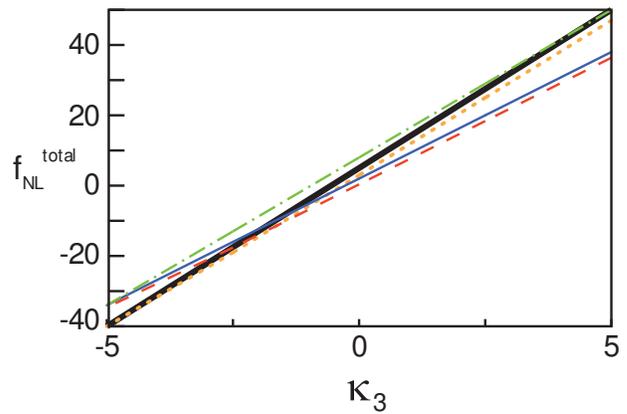}
\caption{\label{Figure4} {\small A comparison of the results from
numerical calculations with the fitting formula given
in Eq.~(\ref{fNLapproximation}) and indicated by the thick black line.
Here we have fixed the value of $\epsilon_{ek} = 36.$
The plot confirms that $f_{NL}$ then grows linearly with $\k_3,$ in good agreement with (\ref{fNLapproximation}). The
sample models (dashed and dotted lines) are representative of the range of models and
parameters shown in \cite{Lehners:2007wc}. We have similarly
checked the dependence on $\sqrt{\e}$ when the value of $\k_3$ is kept fixed.}}
\end{center}
\end{figure}

Putting everything together, we get the following estimates:
\bea f_{NL}^{intrinsic} &\approx& 10\tilde{c} + 50 \\
f_{NL}^{reflection} &\approx& -50 \label{fNLrefestimate} \\
f_{NL}^{integrated} &\approx& 5. \eea The analytic estimate for
$f_{NL}^{reflection}$ suggests a range that could extend to
$-100$, but the value above is more consistent with the exact
numerical calculations for a wide range of parameters.
Altogether, we end up with the following fitting
formula \bea f_{NL}^{total} &\approx&  10 \tilde{c} +5 \\
&\approx& \frac{3}{2}\k_3 \sqrt{\epsilon_{ek}} +5.
\label{fNLapproximation}\eea Fig. \ref{Figure4} shows that this
formula agrees well with the results from exact numerical
calculations. In particular, we have been able to estimate the
slope, and hence the dependence on $\k_3\sqrt{\e_{ek}},$ rather
accurately.

\subsection{Conversion during the ekpyrotic phase} \label{subsection
duringekpyrosis}

In the ``new ekpyrotic'' scenario
\cite{Buchbinder:2007ad,Creminelli:2007aq,Koyama:2007mg}, the
conversion of entropy to curvature perturbations takes place
because the background trajectory switches from the two-field
unstable scaling solution to a single-field attractor solution;
in other words, the trajectory starts out close to the ridge of
the two-field potential and then falls off one of the steep
sides, either because of the initial conditions or due to an
additional feature in the potential. Adding a feature to the
potential makes little difference to the results as long as the
reflection remains gradual, so in fact we will not consider an
additional potential $V^R$ in this section. Since the
conversion happens during the ekpyrotic phase, {\it i.e.} while
the ekpyrotic potentials are relevant,
$\epsilon_c=\epsilon_{ek},$ and from (\ref{fNLestimate}) we
expect the non-gaussianity to be of order \be f_{NL} \sim
\epsilon_{ek} \sim {\cal{O}}(c_1^2). \ee As shown by Koyama
{\it et al.}\cite{Koyama:2007if}, the $\delta N$ formalism is
well suited to treating this case, with the result that the
total local non-linearity parameter is given by \be f_{NL} =
-\frac{5}{12}c_j^2, \label{fNLKoyama}\ee where the index $j$
corresponds to the field that becomes frozen in the late-time
single-field solution (here we have reverted to the potential
form (\ref{c1}), as it is better suited for studying conversion
during the ekpyrotic phase). As a check, we have performed the
direct integration of the equations of motion numerically
\footnote{In their paper, Koyama {\it et al.} also performed a
numerical check within the framework of the $\delta N$
formalism.}, and we find results that are in good agreement
with the above analytic estimate. Note that the sign of
$f_{NL}$ is always negative for these cases.

A qualitative understanding of this result can be achieved in
the present context as follows: from (\ref{fNLintrinsic}) we
can see that, since $H/\dot{\sigma}$ is negative and
approximately constant, the sign of $f_{NL}^{intrinsic}$ is
given by the sign of $\int \dot{\theta} \delta s^{(2)}.$ In
practice, numerical simulations indicate that, for conversion
during the ekpyrotic phase, we have \be |\dot{\theta}| \approx
\frac{1}{10|t|} \label{thetadotekpyrotic} \ee during most of
the conversion, but with $\dot{\theta}$ growing faster towards
the end. We will use this numerical input to guide our
analysis. The sign of $f_{NL}^{intrinsic}$ is essentially
determined by the sign of $\dot{\theta} \delta s^{(2)}$ towards
the end of the period of conversion. Naively it is difficult to
perform a purely analytic estimate of $f_{NL}$ in the current
scheme, since all the terms in the equation of motion
(\ref{eq-entropy}) go as $c_1 t^{-4};$ however, equation
(\ref{thetadotekpyrotic}) tells us that \be |V_{ss}| = 2/t^2
\gg \dot{\theta}^2, \label{Vssdominant}\ee and so there are
surprisingly few terms in the equation of motion for $\delta
s^{(2)}$ that are actually important during (most of) the time
of conversion. In fact, to a first approximation, we are simply
left with \be \ddot{\delta s^{(2)}} = (-\frac{1}{2}V_{sss} +
\frac{\dot{\theta}}{\dot{\sigma}} V_{ss}) (\delta s^{(1)})^2.
\ee $V_{sss}$ decreases in importance as the single-field
scaling solution is reached. Thus, even though the $V_{sss}$
term determines the initial evolution of $\delta s^{(2)},$
eventually the $V_{ss}$ term dominates. Then, since $V_{ss}<0$
it is easy to see that the sign of $\delta s^{(2)}$ is always
driven to be opposite to that of $\dot{\theta},$ and
consequently $f_{NL}^{intrinsic}$ is negative in all cases.

Since $f_{NL}^{reflection}$ is proportional to
${\cal{R}}_L^{-2},$ it provides a contribution of the same order
${\cal{O}}(\epsilon_{ek}).$ Using (\ref{Vssdominant}), it is
straightforward to see that the part of $f_{NL}^{reflection}$
that is proportional to $(\d s)^2$ is always positive, while
the part proportional to $\d s \dot{\d s}$ is always negative.
This implies that there will be a competition between the two
contributions. Numerical integration then shows that the part
proportional to $(\d s)^2$ approximately cancels out
$f_{NL}^{intrinsic},$ while the $\d s \dot{\d s}$ part by
itself is very close in numerical value to the final answer.
In all cases $f_{NL}^{integrated}$ is completely negligible.

\begin{table}
\begin{tabular}{|c|c||c||c|}
  \hline
  $c_1$ & $c_2$ & $f_{NL,\delta N}$ & $f_{NL}$ \\ \hline
  10 & 10 & -41.67  & -39.95   \\
  10 & 15 & -41.67  & -40.45  \\
  10 & 20 & -41.67  & -40.62 \\
  15 & 10 & -93.75  & -91.01 \\
  15 & 15 & -93.75  & -92.11  \\
  15 & 20 & -93.75  & -92.49   \\
  20 & 10 & -166.7  & -162.5   \\
  20 & 15 & -166.7  & -164.4  \\
  20 & 20 & -166.7  & -165.1  \\
  \hline
\end{tabular}
\caption{\small Ekpyrotic conversion: the values of $f_{NL}$
estimated by the $\d N$ formalism compared to the numerical
results obtained by directly integrating the equations of
motion.} \label{Table1}
\end{table}

Consequently, the total $f_{NL}$ turns out to be negative and
moreover in good agreement with the $\delta N$ result
(\ref{fNLKoyama}) \cite{Koyama:2007if}, as shown in Table
\ref{Table1}. Clearly, the $\delta N$ formalism provides a fast
and elegant derivation of non-gaussianity for conversion during
the ekpyrotic phase that agrees well with direct integration of
the equations of motion.  (Note, however, no analogous $\delta
N$ approach applies to conversion in the kinetic energy
dominated phase.)

The conclusion that $f_{NL}$ is substantially less than zero
differs from Ref. \cite{Buchbinder:2007at}; their approximation
method was incomplete though, since the term proportional to
$\d s \dot{\d s}$ in (\ref{zetadotquadratic}) was not included.
As shown above, this term typically contributes significantly
to the final result. Also, the evolution of the entropy
perturbation during conversion was neglected, which meant, for
example, that values of $f_{NL}^{intrinsic}$ of either sign
were obtained. Our results here show that their approximations
were too crude, and that $f_{NL}$ is generally negative with
$f_{NL} \lesssim -20$. This is inconsistent with current limits
by roughly 3 $\sigma$ \cite{Komatsu:2008hk}.

{\it Hence,  we conclude that models with conversion during the
ekpyrotic phase ($\epsilon_c = \epsilon_{ek} \gg 1$) are
difficult to accommodate with current observations in contrast
to models in which conversion occurs during the kinetic energy
dominated phase ($\epsilon_c=3$).}

\subsection{Conversion after the crunch/bang transition}
\label{subsection modulated}

It has recently been proposed by Battefeld
\cite{Battefeld:2007st} that, instead of converting entropy
perturbations into curvature perturbations before the big
crunch/big bang transition, the
 conversion could occur during the phase shortly
following the bang through modulated reheating. The concept is
that massive matter fields are produced copiously at the brane
collision and dominate the energy density immediately after the
bang.  The massive fields are assumed to couple to ordinary
matter with a strength proportional to $h(\delta s)$, so that
their decay into ordinary matter occurs at slightly different
times depending on the value of $\delta s.$  In this way, the
ordinary matter perturbations inherit the entropic perturbation
spectrum. Since the conversion happens while $\epsilon_c
\approx 3,$ we can expect an intrinsic contribution to $f_{NL}$
of \be f_{NL} \sim \sqrt{\epsilon_{ek}} \sim {\cal{O}}(c_1),
\ee {\it i.e.} $f_{NL}$ is of the same order as in the case of
conversion during the kinetic phase preceding the big crunch.
Moreover, as shown in \cite{Battefeld:2007st}, no significant
additional contributions to $f_{NL}$ are expected. Therefore,
models of this type are subject to roughly the same
observational constraints as the models where the conversion
happens during the kinetic phase before the bang and where the
intrinsic contribution to the non-gaussianity is dominant.

Note however that a detailed prediction is made difficult by
the fact that $h$ is an unknown function of $\delta s.$ Indeed,
the entropy perturbations are converted with an efficiency
\cite{Battefeld:2007st} \be e = \frac{3}{2} \frac{|h_{,s}|}{h}.
\ee Since the non-linearity in the entropy field is of
magnitude $\tilde{c},$ this implies a non-linearity in the
curvature perturbation given by \be f_{NL} \approx \pm
\frac{\tilde{c}}{e}, \ee where the $\pm$ sign reflects our
ignorance of the sign of $h_{,s}$ and of the direction of
bending of the scalar field trajectory. Thus, in the absence of
a more detailed model, we cannot go beyond this rough
order-of-magnitude estimate at present.

\section{Conclusions} \label{section conclusion}

Our analysis can be summarized in a few rules of thumb:
\begin{itemize}
\item The intrinsic contribution to $|f_{NL}|$ is
    proportional to the geometric mean of $\epsilon_{ek}$
    and $\epsilon_c$, which is at least two orders of
    magnitude greater than in simple inflationary models
    (where $|f_{NL}^{intrinsic}| = {\cal O}(0.1)$). When
    all contributions are considered, the total $f_{NL}$ is
    generically more than an order of magnitude greater
    than in simple inflationary models, (where
    $f_{NL}^{total} = {\cal O}(1)$).
\item  For a fixed value of $\k_3,$ the value of
    $|f_{NL}^{total}|$ is correlated with the spectral tilt: smaller
    $|f_{NL}^{total}|$ implies smaller $\epsilon_{ek}$, which tends to
    make the spectral tilt bluer. Current limits on $f_{NL}$ fit well
    with limits on the spectral tilt for the simplest models with
    conversion during the kinetic energy driven phase before the bang
    or during reheating after the big bang.
\item Models in which the conversion occurs during a phase
    with larger $\epsilon_c $ produce a  larger intrinsic
    $|f_{NL}|$ and are more difficult to fit with the
    current observed limits on $f_{NL}$ and spectral tilt. In particular, models with conversion in the ekpyrotic phase favor $|f_{NL}|$ large and {\it negative}.
\item Cases in which the intrinsic contribution to $f_{NL}$
    is much smaller than the reflection plus integrated
    contributions produces very large values of $|f_{NL}|
    \gg 100$ that are inconsistent with current
    observational bounds. This includes all cases where the
    conversion is sharp.
\end{itemize}

\begin{figure}[t]
\begin{center}
\includegraphics[width=0.45\textwidth]{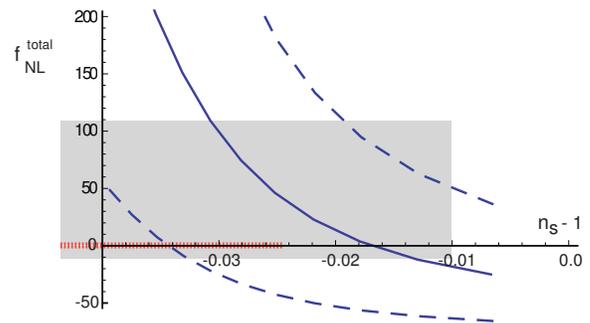}
\caption{\label{fnlvtilt} {\small
A plot for characterizing the correlation
between $|f_{NL}|$ and scalar spectral tilt, $n_s -1$,
here illustrated for the case of  the cyclic model
 in which the conversion from entropic to
curvature perturbations occurs during the kinetic energy dominated
phase just before the big crunch/big bang transition. Different curves correspond to different fixed amounts of skewness $\k_3$ in the potential (the central curve corresponds to $\k_3=4\sqrt{2/3}$), while we vary the steepness of the potential $\epsilon_{ek}.$
The curves  show
the general trend that $|f_{NL}|$ increases as the spectrum becomes redder.
 Simple inflationary models correspond to the narrow
 horizontal hashed (red) strip  with $|f_{NL}| \lesssim 1$.
 The shaded rectangle represents the current observational constraints
 on $f_{NL}$ and tilt (95\% confidence) from WMAP5 \cite{Komatsu:2008hk}.
 }}
\end{center}
\end{figure}

The analysis suggests  a  useful characteristic plot
 for differentiating
cosmological models: $f_{NL}$ versus tilt.
Figure~\ref{fnlvtilt} illustrates the prediction for a cyclic
model in which the conversion from
entropic to curvature perturbations occurs in the kinetic energy
dominated epoch following the ekpyrotic phase. Here we keep the skewness $\k_3$ of the potential fixed, as we vary its steepness $\epsilon.$ The prediction is
a swath whose width is largely due to the uncertainty in the trajectory,
parameterized by $\gamma$. Although the swath includes $f_{NL}$
near zero, positive $f_{NL}$ between 10 and 100 is preferred for tilts
in the range suggested by WMAP5. The prediction for simple inflationary
models is confined to the narrow
band $\Delta f_{NL} \lesssim 1$ around zero.

The results are surprisingly predictive. If observations of
$f_{NL}$ lie in the range predicted by the intrinsic
contribution of either inflationary or ekpyrotic/cyclic models,
it is reasonable to apply Occam's razor and Bayesian analysis
to favor one cosmological model over the other. Combining with
measurements of the spectral tilt significantly sharpens the
test. The current observational bounds obtained by the WMAP
satellite are still inconclusive, but it is clear from the
estimates presented here that non-gaussianity should be
detected by the Planck satellite if the ekpyrotic/cyclic model
is correct. At the same time, this provides a strong incentive
to further refine other methods of measuring non-gaussianity,
such as looking for evidence in measurements of the large scale
structure of the universe.

\section*{Acknowledgements}

We would like to thank Thorsten Battefeld, Evgeny Buchbinder,
Justin Khoury, Burt Ovrut, Neil Turok and David Wands for
helpful and stimulating discussions, and in particular Kazuya
Koyama for discussions and detailed comparisons with the $\d N$ formalism and
S\'{e}bastien Renaux-Petel for pointing out a sign error in
$\tilde{c}$ in an earlier version of this paper. We would also
like to thank the Perimeter Institute for hospitality while
this work was completed. This work is supported in part by the
US Department of Energy grant DE-FG02-91ER40671.

\end{document}